\documentclass[reprint,twocolumn]{revtex4}
\usepackage[pdftex]{graphicx}

\newcommand\ee{\end{equation}}
\newcommand\be{\begin{equation}}

\begin{document}

\title{The neutrino floor at ultra-low threshold}
\author{Louis E. Strigari}

\affiliation{Mitchell Institute for Fundamental Physics and Astronomy,  Department of Physics and Astronomy, Texas A\&M University, College Station, TX 77845, USA}
 
\begin{abstract}
By lowering their energy threshold direct dark matter searches can reach the neutrino floor with experimental technology now in development. The ${}^7Be$ flux can be detected with $\sim 10$ eV nuclear recoil energy threshold and 50 kg-yr exposure. The $pep$ flux can be detected with $\sim 3$ ton-yr exposure, and the first detection of the CNO flux is possible with similar exposure. The $pp$ flux can be detected with threshold of $\sim$ eV and only $\sim$ kg-yr exposure. These can be the first pure neutral current measurements of the low-energy solar neutrino flux. Measuring this flux is important for low mass dark matter searches and for understanding the solar interior. 
\end{abstract}

\maketitle

\section{Introduction} 
\par Particle dark matter with mass near the weak scale has long provided a compelling and testable cosmological paradigm~\cite{Jungman:1995df}. Direct dark matter searches have especially strong bounds on particles of mass $\sim 10$ GeV - $1$ TeV~\cite{Aprile:2012nq,Akerib:2015rjg}, with expected improvement from various experiments in the near future. In addition to improving sensitivity for dark matter in this mass range, there is ample theoretical and experimental motivation to expand the search window. Of particular interest is dark matter with mass $\lesssim 10$ GeV, below which experimental limits have recently improved~\cite{Agnese:2015nto,Angloher:2015ewa,Armengaud:2016cvl} and there has been renewed theoretical emphasis~\cite{Zurek:2013wia,Essig:2013lka}. 

\par Direct dark matter searches will sooner or later be confronted by the ``neutrino floor," which is due to interactions of astrophysical neutrinos~\cite{Billard:2013qya,Ruppin:2014bra,Dent:2016iht}. In particular, a $\lesssim 10$ GeV particle induces a signal in an energy regime that overlaps with the solar neutrino signal. Indeed a major focus of recent discussion is the ${}^8B$ component of the solar neutrino flux, which mimics an $\sim 6$ GeV dark matter particle in future detectors~\cite{Billard:2013qya}. As detectors further lower their thresholds and become sensitive to even lighter dark matter,  lower energy components of the solar neutrino flux, such as  $pp$, ${}^7Be$, $pep$ and $CNO$, will become detectable. Dark matter searches with thresholds low enough to be sensitive to these solar neutrino flux components will see a ``raised" neutrino floor, corresponding to dark matter with spin-independent cross section $\sim 10^{-45}$ cm$^2$~\cite{Billard:2013qya}. 

\par Identifying the neutrino floor is important not only for low mass dark matter searches, it is independently important for understanding the solar metallicity problem and in searches for new physics~\cite{Billard:2014yka}. Recent modeling suggests a lower abundance of metals in the solar core, i.e. a low-$Z$ Standard Solar Model (SSM)~\cite{Asplund:2009fu}, in comparison to the previously established high-$Z$ SSM~\cite{Grevesse:1998bj}. Though some solar neutrino experiments favor a high-$Z$ SSM, a global analysis of all solar neutrino fluxes remains inconclusive~\cite{Antonelli:2012qu,Robertson:2012ib}. 

\par In this paper we discuss the prospects for reaching the neutrino floor with ultra-low threshold dark matter detectors sensitive to nuclear recoils $\sim$ eV. We calculate the detector mass required to measure the low-energy components of the solar neutrino flux, and discuss the complementarity with existing measurements of solar neutrinos. 

\section{Low energy solar neutrinos} 
\par Four components of the solar neutrino flux have now been directly measured: $p + p \rightarrow {}^{2}H + e^+ + \nu_e$ ($pp$), $p + e^{-} + p \rightarrow {}^{2}H + \nu_e$ ($pep$), ${}^8B \rightarrow {}^8Be^\star + e^{+} + \nu_e$ (${}^8B$), and ${}^7Be + e^{-} \rightarrow  {}^{7}Li+ \nu_e$ (${}^7Be$). The $pp$/$pep$ components provide a direct measure of the solar energy generated from the fusion chain, which accounts for $\sim 99\%$ of the solar energy output. The ${}^7Be$ (${}^8B$) fluxes directly measure the respective contributions from the ppII (ppIII) chains. The spectrum of $pep$ neutrinos is a thermally-broadened line at $1.44$ MeV. There are two thermally-broadened lines associated with ${}^7Be$ neutrinos, one at $0.86$ MeV with $\sim 90\%$ branching fraction and a one at $0.38$ MeV with $\sim 10\%$ branching fraction. 

\begin{figure*}
\centering
\begin{minipage}{.5\textwidth}
  \centering
  \includegraphics[width=.65\linewidth]{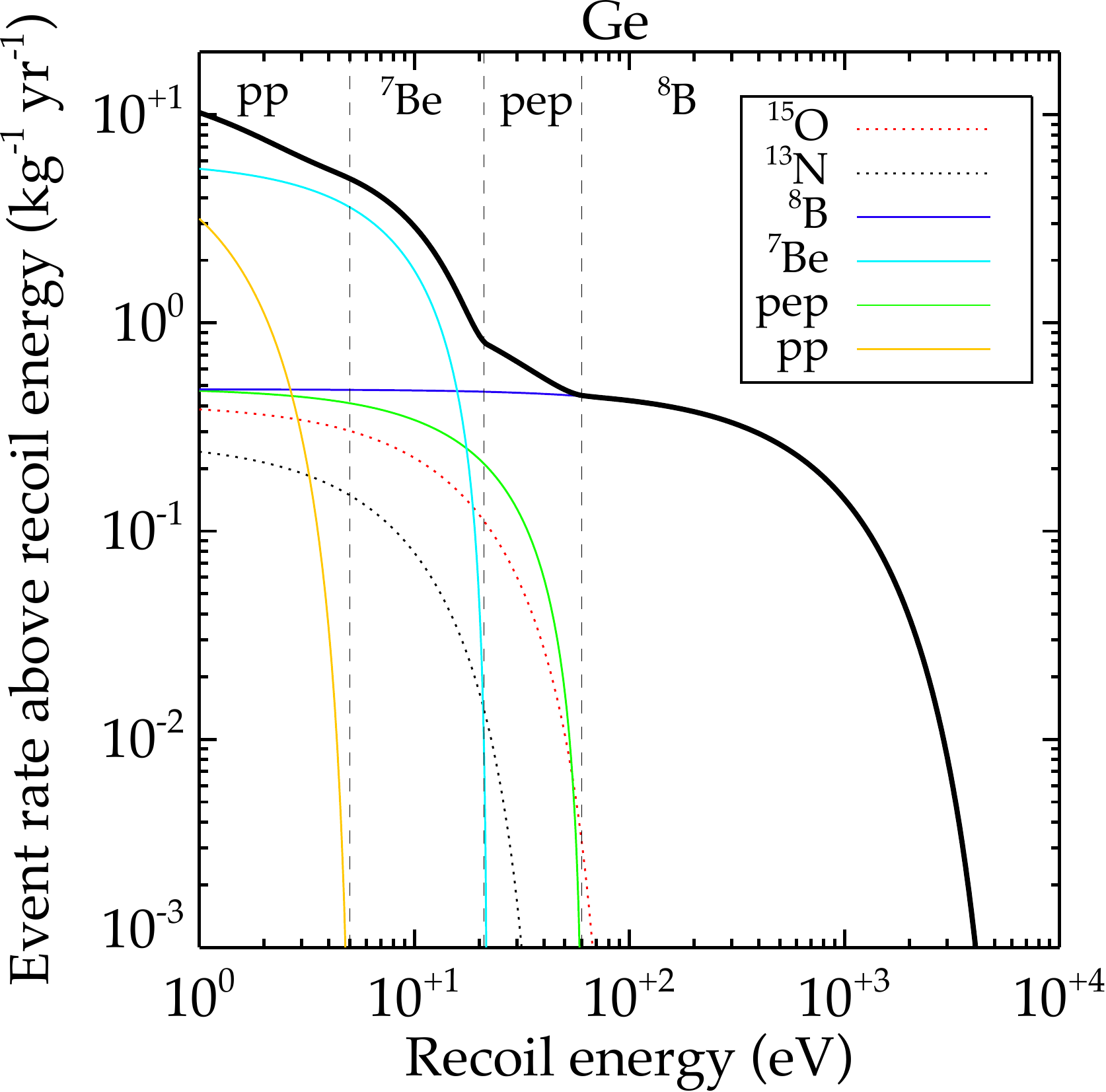}
\end{minipage}%
\begin{minipage}{.5\textwidth}
  \centering
  \includegraphics[width=.65\linewidth]{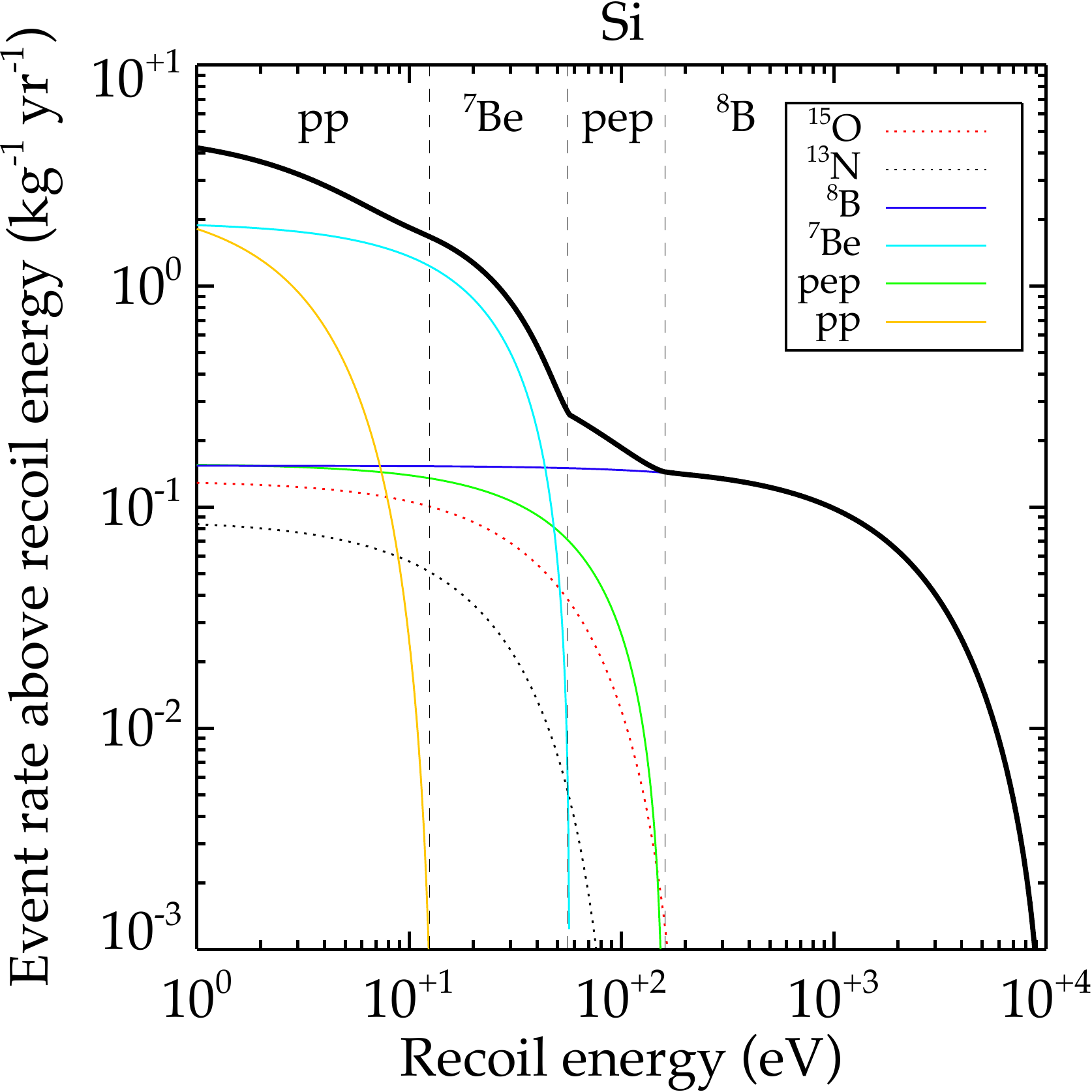}
\end{minipage}
\caption{Integrated number of events above a nuclear recoil threshold energy from coherent scattering due to solar neutrinos in a Ge (left) and Si (right) detector. The black solid curve is the sum of all components. The vertical dashed lines indicate the energy regions over which a particular flux component is dominant.}
\label{fig:spectrum}
\end{figure*}

\par The rate of elastic neutrino-electron interactions from the 0.86 MeV ${}^7Be$ line was measured by the solar neutrino spectroscopy experiment Borexino~\cite{Bellini:2011rx,Bellini:2013lnn}. The equivalent $\nu_e$-electron flux, which is calculated assuming that the observed rate is due only to electron neutrino interactions, is $(2.79 \pm 0.13) \times 10^9$ cm$^{-2}$ s$^{-1}$~\cite{Bellini:2013lnn}. Assuming the MSW solution for $\nu_e$ transition to other flavors and the high-$Z$ SSM, this corresponds to a survival probability of $P_{ee} = 0.51 \pm 0.07$, and the deduced SSM flux is $(4.43 \pm 0.22) \times 10^9$ cm$^{-2}$ s$^{-1}$. This measurement is consistent with the high-$Z$ SSM, which predicts a 0.86 MeV ${}^{7}Be$ flux of $4.47(1 \pm 0.07) \times 10^9$ cm$^{-2}$ s$^{-1}$~\cite{Grevesse:1998bj}, and disfavors the low-$Z$ prediction of $4.08(1 \pm 0.07) \times 10^9$ cm$^{-2}$ s$^{-1}$~\cite{Asplund:2009fu}. The uncertainties on these theoretical fluxes are due to variations of the SSM parameters. 

\par Borexino has also measured the flux from the $pep$ reaction, again from elastic neutrino-electron interactions~\cite{Collaboration:2011nga,Bellini:2013lnn}. For $pep$ neutrinos the equivalent $\nu_e$-electron flux is $(1.00 \pm 0.22) \times 10^8$ cm$^{-2}$ s$^{-1}$. Assuming the MSW solution for $\nu_e$ transition to other flavors,  the survival probability at 1.44 MeV is $P_{ee} = 0.62 \pm 0.17$, and the deduced SSM flux is  $(1.63 \pm 0.35) \times 10^8$ cm$^{-2}$ s$^{-1}$. The $pep$ flux measured by Borexino is consistent with both the high and low-$Z$ SSMs, though the predicted $pep$ flux is relatively insensitive to solar metallicity. Because the $pep$ flux is closely related to the solar luminosity and to the $pp$ flux, there is a low theoretical uncertainty on the $pep$ flux of $\sim 0.01$. 

\par Borexino has recently reported the first direct measurement of the $pp$ neutrino spectrum~\cite{Bellini:2014uqa}. Assuming the MSW solution, the deduced SSM flux is $(6.6 \pm 0.7) \times 10^{10}$ cm$^{-2}$ s$^{-1}$. This is in agreement with the predictions of both the high-$Z$ ($5.98 \times 10^{10}$ cm$^{-2}$ s$^{-1}$) and low-$Z$ SSM ($6.03 \times 10^{10}$ cm$^{-2}$ s$^{-1}$). Again the small variation in these predictions is because the $pp$ flux is directly related to the solar luminosity and has a theoretical uncertainty of only $\sim 0.006$. 

\par In addition to the ${}^7Be$, $pep$, and $pp$ measurements, Borexino has placed an upper bound of $< 7.7 \times 10^8$ cm$^{-2}$ s$^{-1}$ on the sum of all components that contribute to the $CNO$ flux. This corresponds to a ratio of the flux with respect to the high-$Z$ SSM prediction of $< 1.5$. There is a relatively large theoretical uncertainty ($\sim 0.15$) on the $CNO$ neutrino flux, and it is very sensitive to the solar metallicity. 

\begin{figure*}
\begin{tabular}{cccc}
\includegraphics[height=4.3cm]{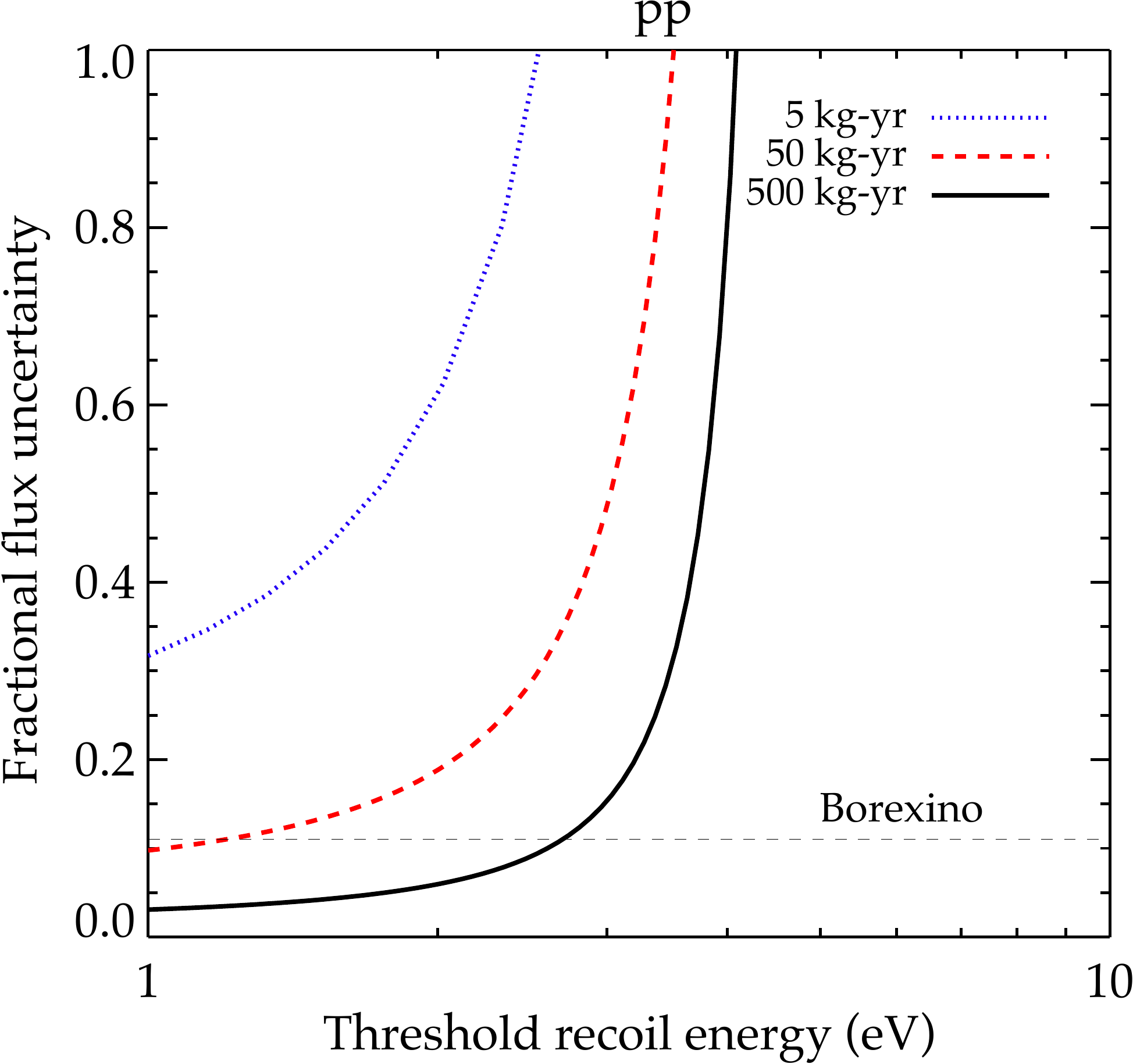} &
\includegraphics[height=4.3cm]{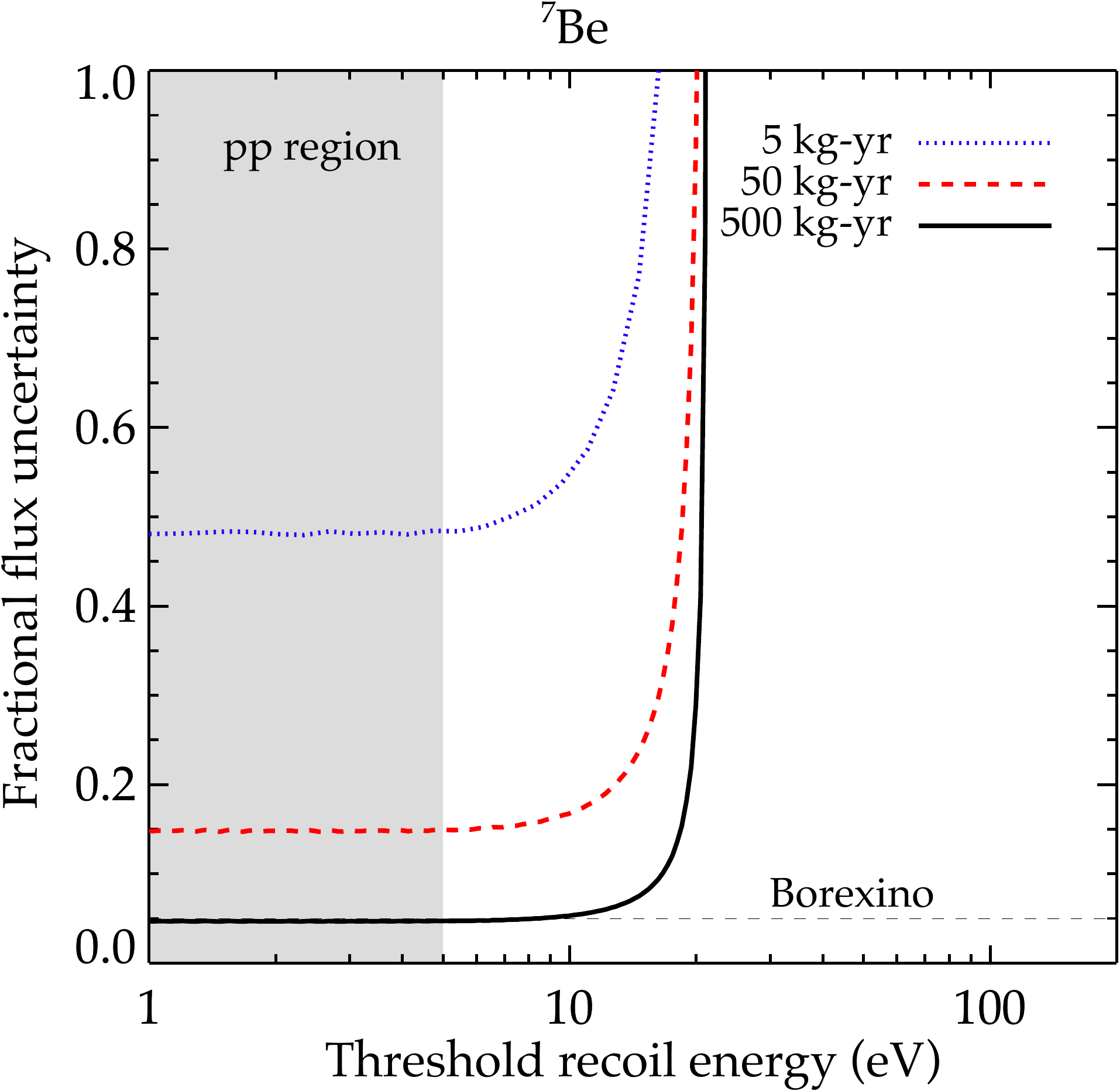} & 
\includegraphics[height=4.3cm]{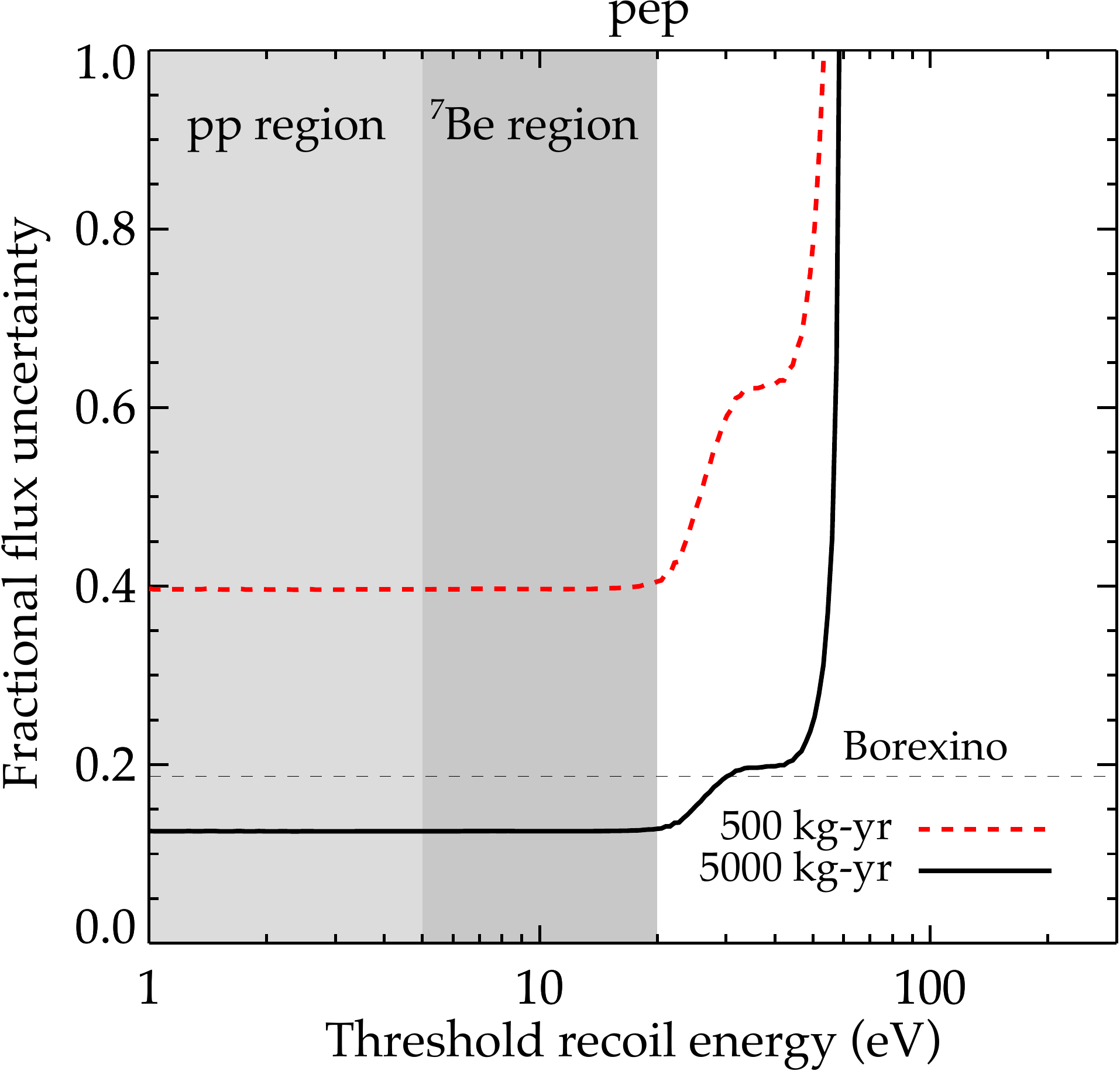} &
\includegraphics[height=4.3cm]{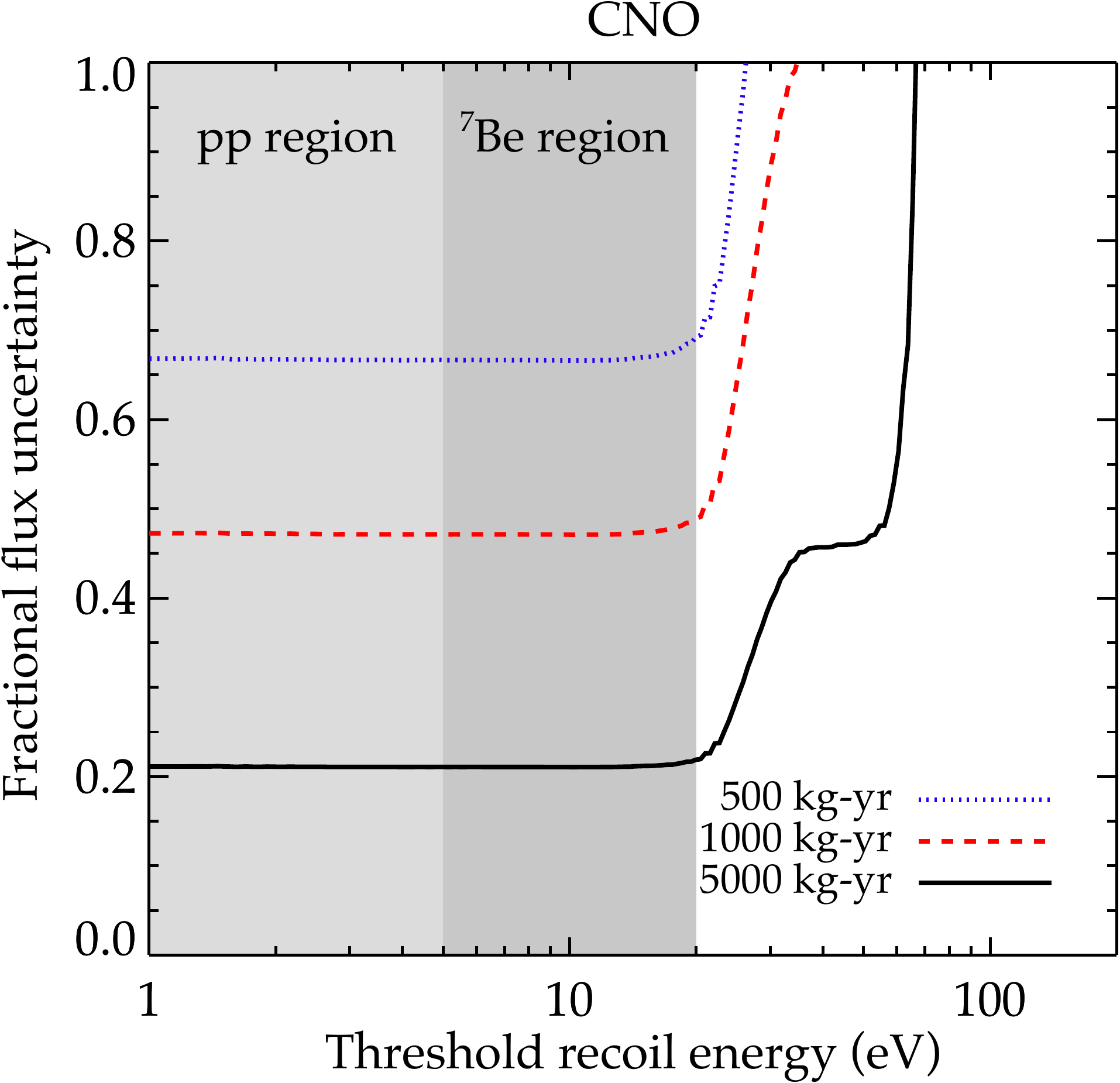} \\
\includegraphics[height=4.3cm]{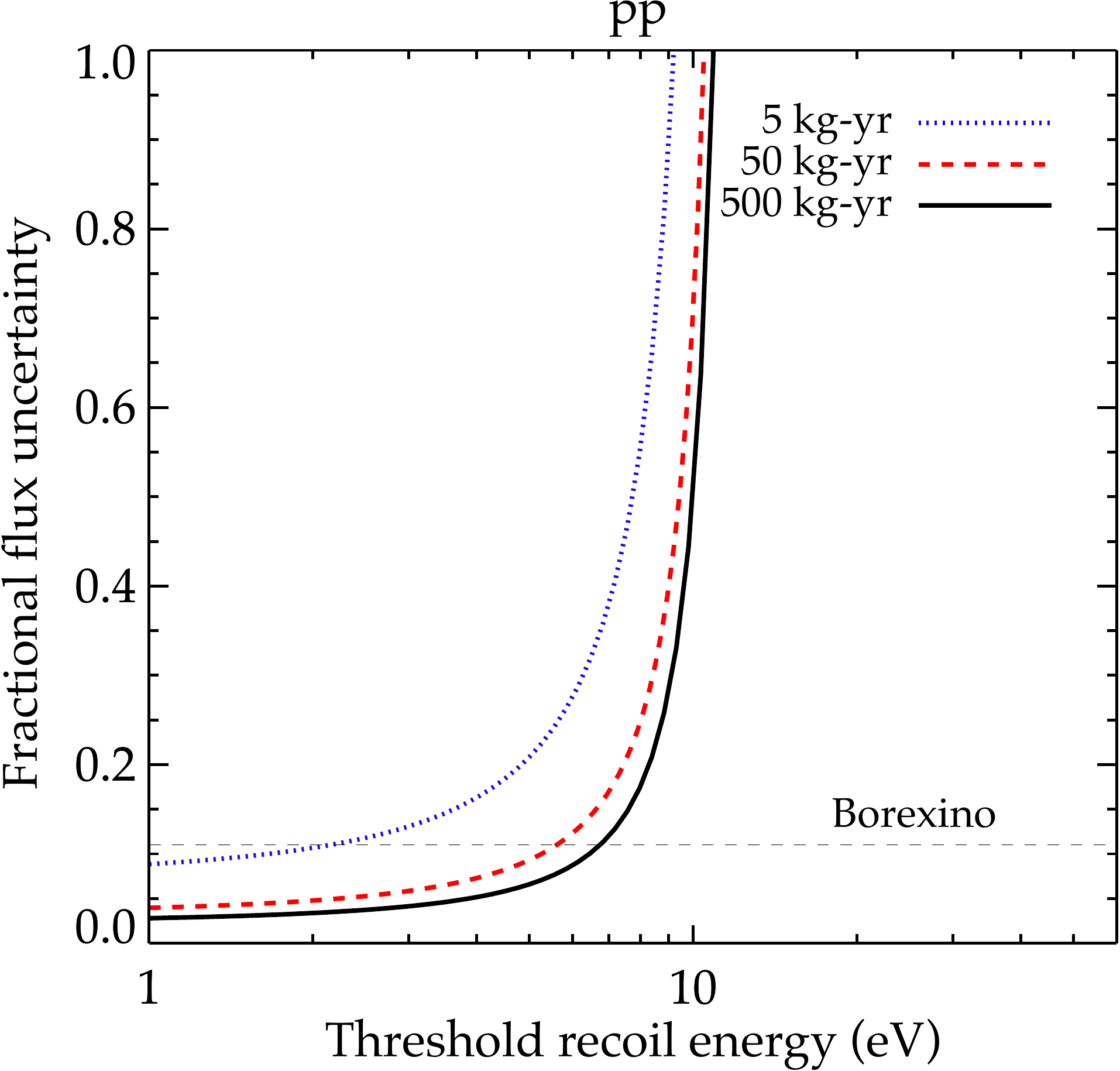} &
\includegraphics[height=4.3cm]{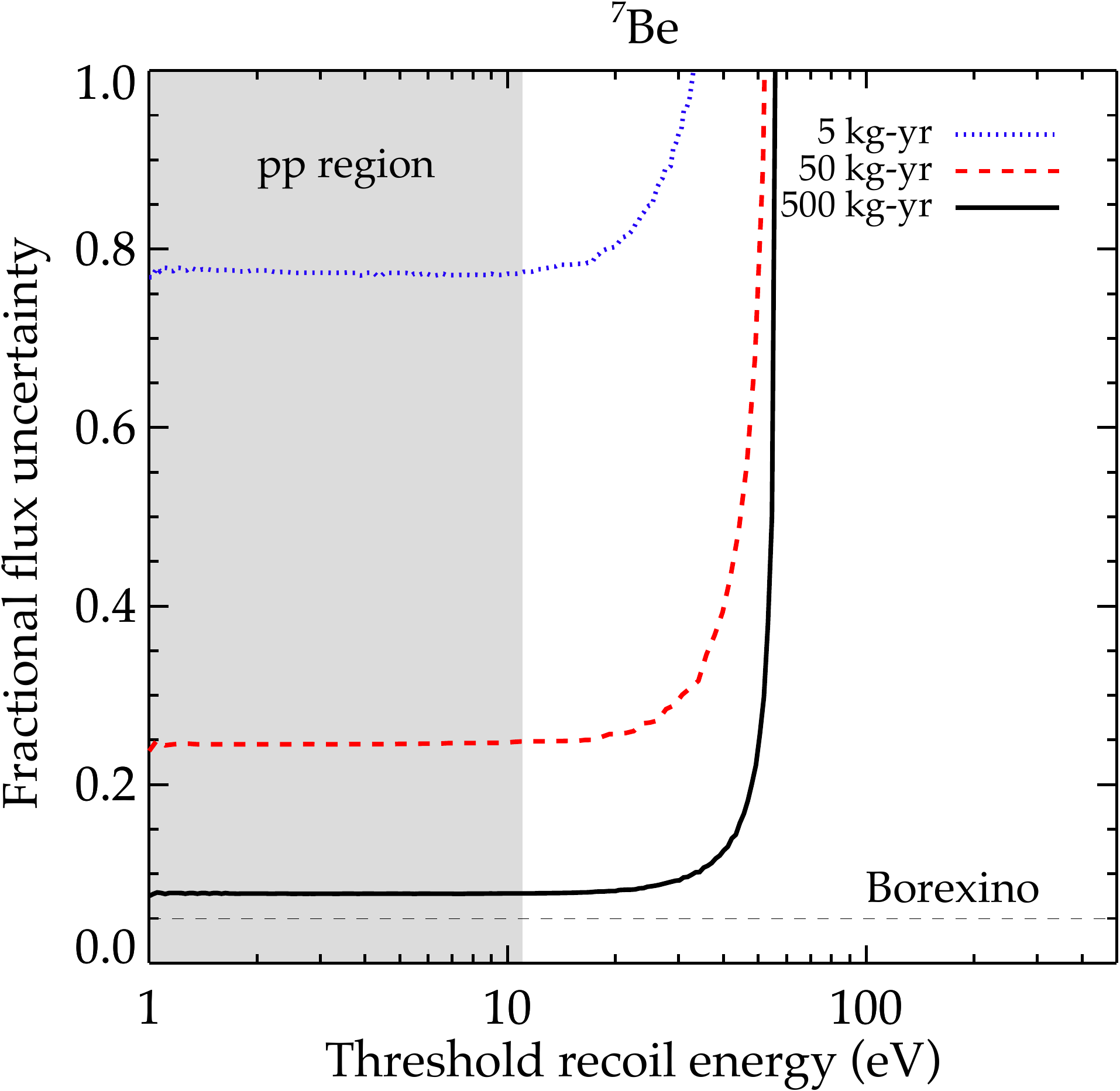} & 
\includegraphics[height=4.3cm]{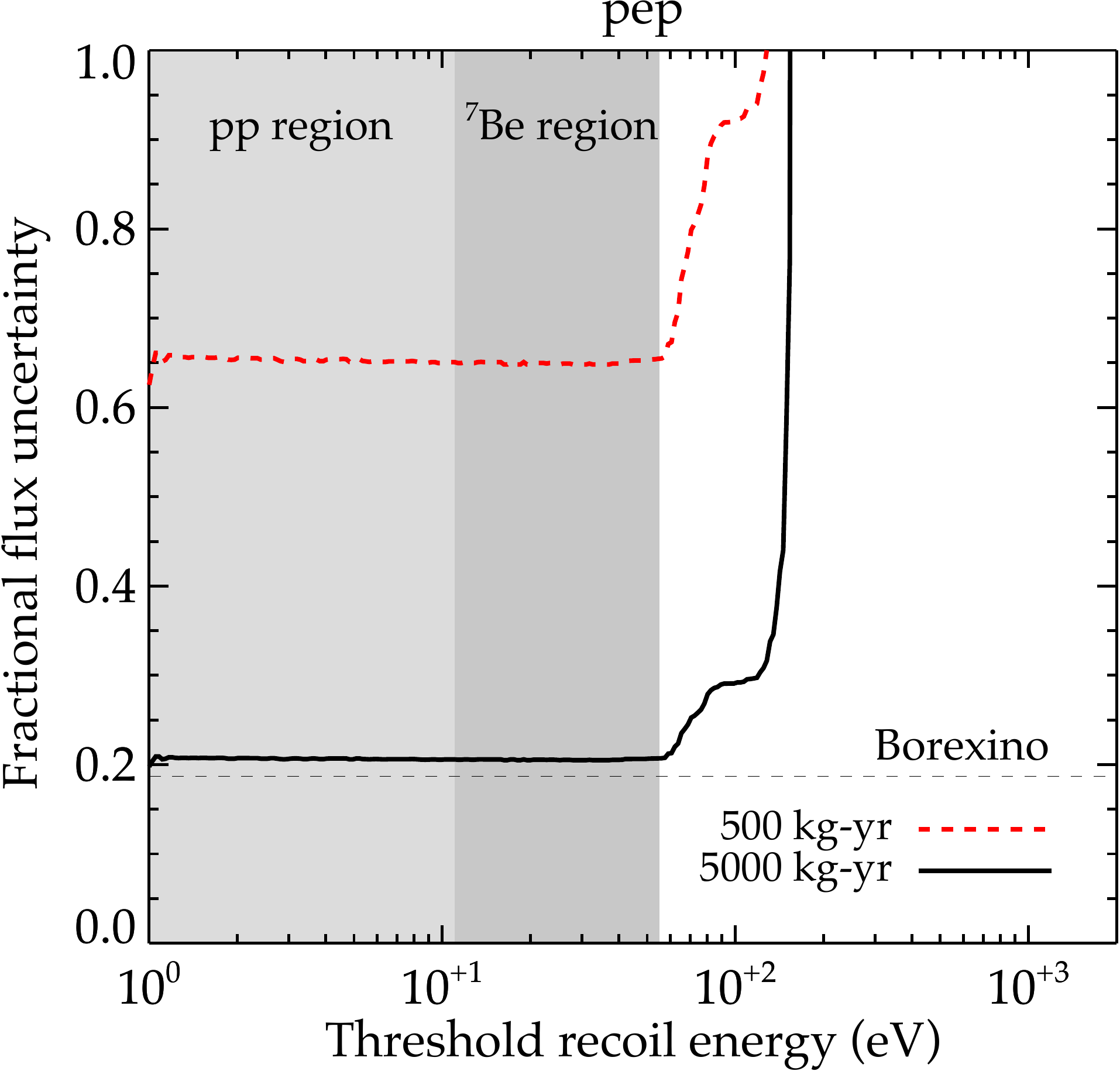} &
\includegraphics[height=4.3cm]{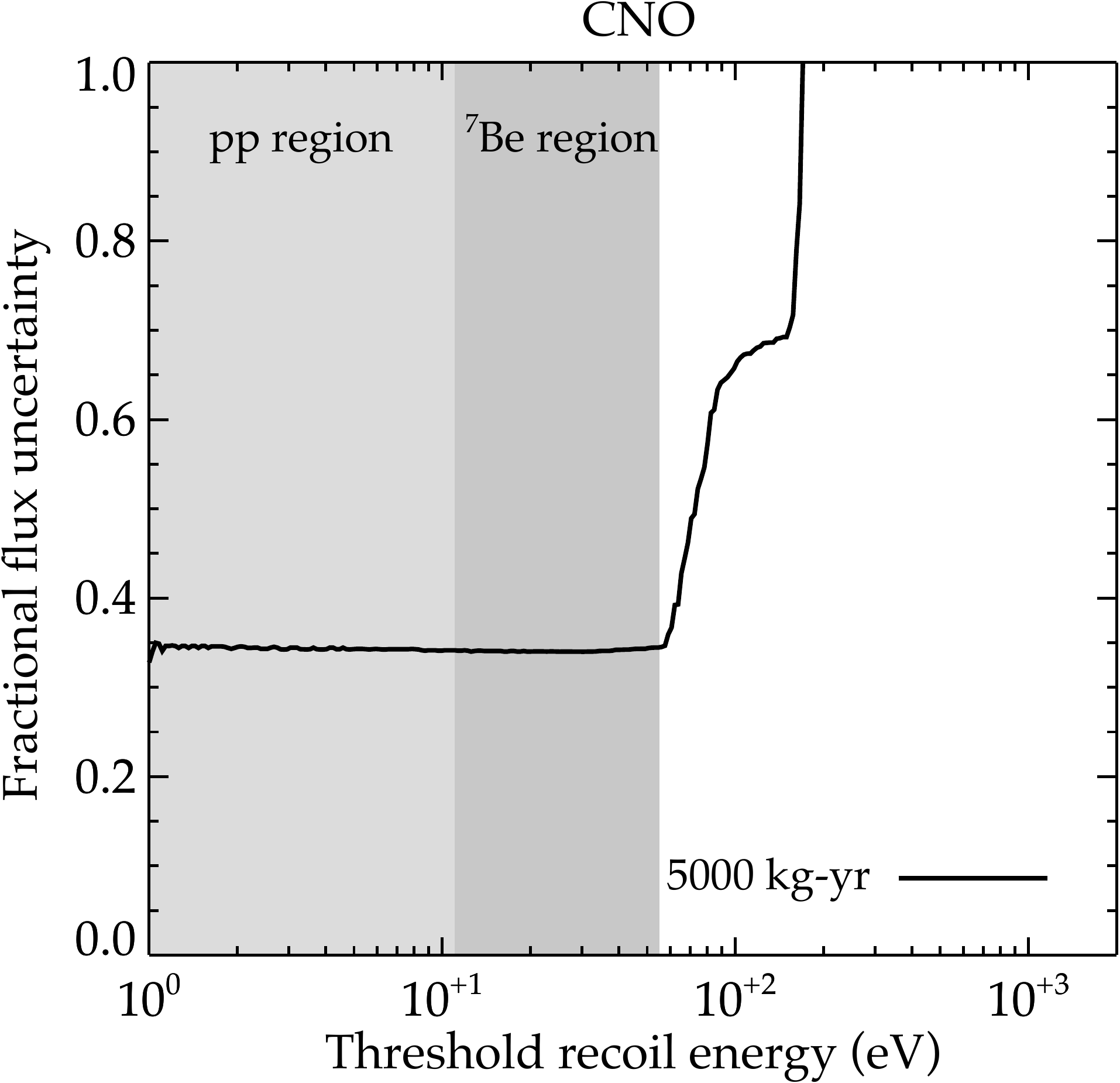} \\
\end{tabular}
\caption{Fractional flux uncertainty ($\Delta f$) on the $pp$, ${}^7Be$, $pep$, and $CNO$ components as a function of threshold nuclear recoil energy. The top row is for Ge, and the bottom row is for Si. For the $pp$, ${}^7Be$, and $pep$ panels, the Borexino sensitivity is indicated. In the ${}^7Be$, $pep$, and $CNO$ panels, energy regions where the $pp$ signal dominates is shaded light grey. In the $pep$ and $CNO$ panels, energy regions where the $pp$ signal dominates is shaded light grey, and energy regions where the ${}^7Be$ signal dominates is shaded dark grey. Note the difference in energy ranges between the panels, and the different curves in each panel correspond to different exposures.}
\label{fig:fractional_flux_uncertainty}
\end{figure*}

\section{Analysis}
\par In this paper we are interested in detecting solar neutrinos via coherent neutrino-nucleus scattering (CNS) using direct dark matter detection experiments. Because it is a neutral current interaction, the detection of CNS would provide the first direct measurement of the SSM flux and thus a direct measurement of the survival probability for the low-energy solar neutrino spectrum. (Recall that the SNO experiment was sensitive to neutral current interactions of the high-energy ${}^{8}B$ flux~\cite{Aharmim:2011vm}.) We assume pure Standard Model (SM) contributions to the CNS cross section (see Refs.~\cite{Barranco:2005yy,Dutta:2015vwa} for discussions of beyond the SM contributions to the cross section). We further assume the standard three neutrino flavors, which implies that we do not need to account for flavor mixing. We specifically consider ultra-low threshold Ge and Si detectors, whose feasibility has been recently discussed elsewhere~\cite{Mirabolfathi:2015pha}. 

\par Figure~\ref{fig:spectrum} shows the nuclear recoil energy spectrum from solar neutrinos, highlighting the ultra-low threshold regime down to $\sim$ eV. Going from high to low nuclear recoil energy threshold, the event rate is dominated by ${}^8B$, then $pep$, ${}^7Be$, and finally $pp$ solar neutrinos. The most prominent rise in the event rate occurs when the ${}^7Be$ window opens up, corresponding to a nuclear recoil energy threshold of $\sim 20$ eV in Ge, and $\sim 50$ eV in Si. Smaller increases arise in the transition from the ${}^8B$ to $pep$-dominated recoil energy region, and the ${}^7Be$ to the $pp$-dominated recoil energy region. The $CNO$ flux, which we take as the sum of the ${}^{15}O$ and ${}^{13}N$ components, contributes as a subdominant component in an energy region overlapping with $pep$ neutrinos. (Note that we do not consider the ${}^{17}F$ component of the $CNO$ flux, which makes a negligible contribution to the event rate.) 

\par Since several flux components contribute to the energy regions in Figure~\ref{fig:spectrum}, a multi-component spectral analysis is required to detect a particular flux component. To determine the detection prospects, we define a poisson likelihood function in the recoil energy bins and have 
\be
F_{\alpha \beta} = \sum_{\imath=1}^n \frac{T_{exp}^2}{N_\imath^{tot}}  
         \frac{N_{\alpha \imath}}{f_\alpha} 
         \frac{N_{\beta \imath}}{f_\beta}, 
\label{eq:likelihood} 
\ee 
where the sum is over $n$ recoil energy bins, $N_{\alpha \imath}$ is the predicted rate in the $\imath^{th}$ energy bin from a flux component, and $f_\alpha$ are the flux normalizations for each component of the solar neutrino spectrum, so that in our case $\alpha = pp, {}^7Be, \, pep, \, CNO \, \textrm{and} \, {}^8B$. The total number of events from all flux components is $N_\imath^{tot}$, and the exposure, $T_{exp}$, is the mass of the detector times the run time of the experiment. The one-sigma uncertainty on the flux normalization $f_\alpha$ is $\sqrt{ ({\bf F}^{-1})_{\alpha \alpha}}$. 

\par Motivated by developing detector technology with excellent energy resolution~\cite{Mirabolfathi:2015pha}, we examine the event rate in nuclear recoil energy bins of width $\sim$ eV. We quote results in terms of the fractional uncertainty on the flux normalization, $\Delta f$, and quantify how the measurement of $\Delta f$ for each component improves with decreasing nuclear recoil energy threshold and increasing exposure. 

\par For our fiducial model we assume the high-$Z$ SSM for the flux normalizations. Figure~\ref{fig:fractional_flux_uncertainty} shows $\Delta f$ for the $pp$, ${}^7Be$, $pep$, $CNO$ fluxes as a function of threshold nuclear recoil energy for different exposures $T_{exp}$. In all cases there is a dramatic improvement in the measurement of $\Delta f$ as the threshold is dropped into the regimes where each respective flux component dominates (Figure~\ref{fig:spectrum}). For $pp$ neutrinos, a Si detector reaches the Borexino sensitivity for a threshold $\lesssim 3$ eV and an exposure $\sim 5$ kg-yr, while a Ge detector reaches the Borexino sensitivity for the same threshold and an exposure $\sim 500$ kg-yr. It should be emphasized that the Borexino measurement is neutrino-electron scattering, which is due mostly to charged-current interactions. A CNS measurement would thus represent the first pure neutral current detection of these flux components. 

\par For the ${}^7Be$ flux, a $\sim 50$ kg-yr Ge exposure with $\sim 10$ eV threshold will result in a detection with $\Delta f \simeq 0.15$. At this same threshold, $\sim 500$ kg-yr exposure with Ge will match the Borexino sensitivity, $\Delta f \simeq 0.05$. For Si, $\sim 50$ kg-yr exposure with a $\sim 30$ eV threshold will result in a detection with $\Delta f \simeq 0.25$, and a $\gtrsim 500$ kg-yr exposure matches the Borexino sensitivity. Thus for $\gtrsim 1$ eV threshold, a Si detector is most sensitive to the $pp$ flux, while a Ge detector is most sensitive to the ${}^7Be$ flux. 

\par The $pep$ and $CNO$ fluxes are prominent at energies lower than ${}^8B$, but higher than ${}^7Be$. Though the $pep$ and $CNO$ spectral shapes are different, their flux normalizations are correlated in a multi-component analysis. This is evident in Figure~\ref{fig:spectrum} which indicates a brief saturation as the threshold is lowered before $\Delta f$ is ultimately minimized. For the $pep$ flux, we find that a $\sim 500$ kg-yr Ge exposure with $\sim 10$ eV threshold will measure normalization to a fractional uncertainty of $\sim 0.4$. This exposure will provide a $\sim 2\sigma$ detection of the $CNO$ flux. Increasing the exposure to $5$ ton-yr will match the Borexino charged current sensitivity to the $pep$ flux, and also attain $\Delta f \sim 0.2$ on the $CNO$ flux. 

\section{Discussion and conclusion}
\par We have examined the potential for direct dark matter searches to reach the neutrino floor with detector mass similar to those under development and with ultra-low energy thresholds, as low as $\sim$ eV. These detectors, such as e.g. SuperCDMS~\cite{Agnese:2015nto}, will be sensitive to dark matter with mass $\sim$ GeV. For reasonable detector mass $\sim 50$ kg-yr, a threshold of  $\sim 10 \, (30)$ eV in Ge (Si) will measure the ${}^{7}Be$ solar neutrino flux. Approximately an order of magnitude larger mass detectors will be sensitive to $pep$ and $CNO$ neutrinos. For a threshold of a few eV, the $pp$ flux can be identified in both Si and Ge. 

\par Identifying and measuring the neutrino floor in direct dark matter searches is of obvious importance for studying low mass dark matter. It also represents an important step in the continuing development of the solar neutrino program, dating back to over half of a century. The detector technology discussed in this paper will establish the first pure neutral current detection of all the low energy components of the solar neutrino flux, which will be the first direct and model independent measurement of the neutrino survival probability in the vacuum-dominated regime. The excellent energy resolution will in addition provide the first measurement of the energy dependence of the survival probability, which can have important implications in searches for new physics. 

\par If the technology discussed here were to reach ton-scale mass, it will establish the first measurement of neutrinos from the $CNO$ cycle. This is a long sought-after component of the solar neutrino spectrum that generates $\sim 1\%$ of solar energy. A measurement of $CNO$ neutrinos will be important for understanding the solar abundance problem and for understanding the fusion process in main-sequence stars more massive than the Sun. Current scintillation detectors such as Borexino are limited in measuring the $CNO$ flux because of muon induced backgrounds, though current designs may improve upon this present situation~\cite{Andringa:2015tza}. 

\par This analysis has focused on the detection of nuclear recoil events. Future direct dark matter searches will also be sensitive to dark matter and neutrino scattering off of electrons. (For ideas to detect electrons with even lower energies than discussed here see Ref.~\cite{Hochberg:2015fth}.) The major sensitivity is to $pp$ neutrinos, for which the integrated event rate above $1$ keV electron recoil energy is $\sim 4$ kg$^{-1}$ yr$^{-1}$, and the rate remains constant down to eV energies. Over this same electron recoil energy region, the rate due to ${}^7Be$ electron scattering events is $\sim 1/3$ that of the $pp$ rate. Thus for electron recoils there is no substantial gain to lower thresholds, unless the neutrino has a magnetic moment much larger than predicted in the SM. 

{\bf Acknowledgements}:
I am grateful for discussions with Rupak Mahapatra. This work is supported by NSF grant PHY-1522717 and the Mitchell Institute for Fundamental Physics and Astronomy at Texas A\&M University.


\begin{thebibliography}{99}
%\cite{Jungman:1995df}
\bibitem{Jungman:1995df} 
  G.~Jungman, M.~Kamionkowski and K.~Griest,
  %``Supersymmetric dark matter,''
  Phys.\ Rept.\  {\bf 267}, 195 (1996)
  %doi:10.1016/0370-1573(95)00058-5
  [hep-ph/9506380].
  %%CITATION = doi:10.1016/0370-1573(95)00058-5;%%
  %2770 citations counted in INSPIRE as of 17 Mar 2016

%\cite{Akerib:2015rjg}
\bibitem{Akerib:2015rjg} 
  D.~S.~Akerib {\it et al.} [LUX Collaboration],
  %``Improved WIMP scattering limits from the LUX experiment,''
  arXiv:1512.03506 [astro-ph.CO].
  %%CITATION = ARXIV:1512.03506;%%
  %32 citations counted in INSPIRE as of 14 Mar 2016
  
  %\cite{Aprile:2012nq}
\bibitem{Aprile:2012nq} 
  E.~Aprile {\it et al.} [XENON100 Collaboration],
  %``Dark Matter Results from 225 Live Days of XENON100 Data,''
  Phys.\ Rev.\ Lett.\  {\bf 109}, 181301 (2012)
  %doi:10.1103/PhysRevLett.109.181301
  [arXiv:1207.5988 [astro-ph.CO]].
  %%CITATION = doi:10.1103/PhysRevLett.109.181301;%%
  %1095 citations counted in INSPIRE as of 17 Mar 2016
  
%\cite{Agnese:2015nto}
\bibitem{Agnese:2015nto} 
  R.~Agnese {\it et al.} [SuperCDMS Collaboration],
  %``New Results from the Search for Low-Mass Weakly Interacting Massive Particles with the CDMS Low Ionization Threshold Experiment,''
  Phys.\ Rev.\ Lett.\  {\bf 116}, no. 7, 071301 (2016)
  %doi:10.1103/PhysRevLett.116.071301
  [arXiv:1509.02448 [astro-ph.CO]].
  %%CITATION = doi:10.1103/PhysRevLett.116.071301;%%
  %19 citations counted in INSPIRE as of 19 Mar 2016  

%\cite{Angloher:2015ewa}
\bibitem{Angloher:2015ewa} 
  G.~Angloher {\it et al.} [CRESST Collaboration],
  %``Results on light dark matter particles with a low-threshold CRESST-II detector,''
  Eur.\ Phys.\ J.\ C {\bf 76}, no. 1, 25 (2016)
  %doi:10.1140/epjc/s10052-016-3877-3
  [arXiv:1509.01515 [astro-ph.CO]].
  %%CITATION = doi:10.1140/epjc/s10052-016-3877-3;%%
  %18 citations counted in INSPIRE as of 19 Mar 2016

%\cite{Armengaud:2016cvl}
\bibitem{Armengaud:2016cvl} 
  E.~Armengaud {\it et al.} [EDELWEISS Collaboration],
  %``Constraints on low-mass WIMPs from the EDELWEISS-III dark matter search,''
  arXiv:1603.05120 [astro-ph.CO].
  %%CITATION = ARXIV:1603.05120;%%
  
  %\cite{Zurek:2013wia}
\bibitem{Zurek:2013wia} 
  K.~M.~Zurek,
  %``Asymmetric Dark Matter: Theories, Signatures, and Constraints,''
  Phys.\ Rept.\  {\bf 537}, 91 (2014)
  %doi:10.1016/j.physrep.2013.12.001
  [arXiv:1308.0338 [hep-ph]].
  %%CITATION = doi:10.1016/j.physrep.2013.12.001;%%
  %145 citations counted in INSPIRE as of 14 Mar 2016
  
%\cite{Essig:2013lka}
\bibitem{Essig:2013lka} 
  R.~Essig {\it et al.},
  %``Working Group Report: New Light Weakly Coupled Particles,''
  arXiv:1311.0029 [hep-ph].
  %%CITATION = ARXIV:1311.0029;%%
  %163 citations counted in INSPIRE as of 16 Mar 2016  

%\cite{Billard:2013qya}
\bibitem{Billard:2013qya} 
  J.~Billard, L.~Strigari and E.~Figueroa-Feliciano,
  %``Implication of neutrino backgrounds on the reach of next generation dark matter direct detection experiments,''
  Phys.\ Rev.\ D {\bf 89}, no. 2, 023524 (2014)
 % doi:10.1103/PhysRevD.89.023524
  [arXiv:1307.5458 [hep-ph]].
  %%CITATION = doi:10.1103/PhysRevD.89.023524;%%
  %109 citations counted in INSPIRE as of 13 Mar 2016
  
  %\cite{Ruppin:2014bra}
\bibitem{Ruppin:2014bra} 
  F.~Ruppin, J.~Billard, E.~Figueroa-Feliciano and L.~Strigari,
  %``Complementarity of dark matter detectors in light of the neutrino background,''
  Phys.\ Rev.\ D {\bf 90}, no. 8, 083510 (2014)
  %doi:10.1103/PhysRevD.90.083510
  [arXiv:1408.3581 [hep-ph]].
  %%CITATION = doi:10.1103/PhysRevD.90.083510;%%
  %19 citations counted in INSPIRE as of 15 Mar 2016
  
  %\cite{Dent:2016iht}
\bibitem{Dent:2016iht} 
  J.~B.~Dent, B.~Dutta, J.~L.~Newstead and L.~E.~Strigari,
  %``No $\nu$ floors: Effective field theory treatment of the neutrino background in direct dark matter detection experiments,''
  arXiv:1602.05300 [hep-ph].
  %%CITATION = ARXIV:1602.05300;%%

    %\cite{Billard:2014yka}
\bibitem{Billard:2014yka} 
  J.~Billard, L.~E.~Strigari and E.~Figueroa-Feliciano,
  %``Solar neutrino physics with low-threshold dark matter detectors,''
  Phys.\ Rev.\ D {\bf 91}, no. 9, 095023 (2015)
  %doi:10.1103/PhysRevD.91.095023
  [arXiv:1409.0050 [astro-ph.CO]].
  %%CITATION = doi:10.1103/PhysRevD.91.095023;%%
  %9 citations counted in INSPIRE as of 15 Mar 2016

  %\cite{Asplund:2009fu}
\bibitem{Asplund:2009fu} 
  M.~Asplund, N.~Grevesse, A.~J.~Sauval and P.~Scott,
  %``The chemical composition of the Sun,''
  Ann.\ Rev.\ Astron.\ Astrophys.\  {\bf 47}, 481 (2009)
 % doi:10.1146/annurev.astro.46.060407.145222
  [arXiv:0909.0948 [astro-ph.SR]].
  %%CITATION = doi:10.1146/annurev.astro.46.060407.145222;%%
  %1167 citations counted in INSPIRE as of 13 Mar 2016  
  
  %\cite{Grevesse:1998bj}
\bibitem{Grevesse:1998bj} 
  N.~Grevesse and A.~J.~Sauval,
  %``Standard Solar Composition,''
  Space Sci.\ Rev.\  {\bf 85}, 161 (1998).
  %doi:10.1023/A:1005161325181
  %%CITATION = doi:10.1023/A:1005161325181;%%
  %1396 citations counted in INSPIRE as of 13 Mar 2016  
  
%\cite{Antonelli:2012qu}
\bibitem{Antonelli:2012qu} 
  V.~Antonelli, L.~Miramonti, C.~Pena Garay and A.~Serenelli,
  %``Solar Neutrinos,''
  Adv.\ High Energy Phys.\  {\bf 2013}, 351926 (2013)
  %doi:10.1155/2013/351926
  [arXiv:1208.1356 [hep-ex]].
  %%CITATION = doi:10.1155/2013/351926;%%
  %23 citations counted in INSPIRE as of 14 Mar 2016

%\cite{Robertson:2012ib}
\bibitem{Robertson:2012ib} 
  W.~C.~Haxton, R.~G.~Hamish Robertson and A.~M.~Serenelli,
  %``Solar Neutrinos: Status and Prospects,''
  Ann.\ Rev.\ Astron.\ Astrophys.\  {\bf 51}, 21 (2013)
 % doi:10.1146/annurev-astro-081811-125539
  [arXiv:1208.5723 [astro-ph.SR]].
  %%CITATION = doi:10.1146/annurev-astro-081811-125539;%%
  %56 citations counted in INSPIRE as of 14 Mar 2016

%\cite{Bellini:2011rx}
\bibitem{Bellini:2011rx} 
  G.~Bellini {\it et al.},
  %``Precision measurement of the 7Be solar neutrino interaction rate in Borexino,''
  Phys.\ Rev.\ Lett.\  {\bf 107}, 141302 (2011)
  %doi:10.1103/PhysRevLett.107.141302
  [arXiv:1104.1816 [hep-ex]].
  %%CITATION = doi:10.1103/PhysRevLett.107.141302;%%
  %234 citations counted in INSPIRE as of 13 Mar 2016

%\cite{Bellini:2013lnn}
\bibitem{Bellini:2013lnn} 
  G.~Bellini {\it et al.} [Borexino Collaboration],
  %``Final results of Borexino Phase-I on low energy solar neutrino spectroscopy,''
  Phys.\ Rev.\ D {\bf 89}, no. 11, 112007 (2014)
  %doi:10.1103/PhysRevD.89.112007
  [arXiv:1308.0443 [hep-ex]].
  %%CITATION = doi:10.1103/PhysRevD.89.112007;%%
  %81 citations counted in INSPIRE as of 13 Mar 2016

  %\cite{Collaboration:2011nga}
\bibitem{Collaboration:2011nga} 
  G.~Bellini {\it et al.} [Borexino Collaboration],
  %``First evidence of pep solar neutrinos by direct detection in Borexino,''
  Phys.\ Rev.\ Lett.\  {\bf 108}, 051302 (2012)
  %doi:10.1103/PhysRevLett.108.051302
  [arXiv:1110.3230 [hep-ex]].
  %%CITATION = doi:10.1103/PhysRevLett.108.051302;%%
  %135 citations counted in INSPIRE as of 13 Mar 2016


 %\cite{Bellini:2014uqa}
\bibitem{Bellini:2014uqa} 
  G.~Bellini {\it et al.} [BOREXINO Collaboration],
  %``Neutrinos from the primary protonÐproton fusion process in the Sun,''
  Nature {\bf 512}, no. 7515, 383 (2014).
 % doi:10.1038/nature13702
  %%CITATION = doi:10.1038/nature13702;%%
  %50 citations counted in INSPIRE as of 16 Mar 2016
    
%\cite{Aharmim:2011vm}
\bibitem{Aharmim:2011vm} 
  B.~Aharmim {\it et al.} [SNO Collaboration],
  %``Combined Analysis of all Three Phases of Solar Neutrino Data from the Sudbury Neutrino Observatory,''
  Phys.\ Rev.\ C {\bf 88}, 025501 (2013)
  doi:10.1103/PhysRevC.88.025501
  [arXiv:1109.0763 [nucl-ex]].
  %%CITATION = doi:10.1103/PhysRevC.88.025501;%%
  %173 citations counted in INSPIRE as of 19 Mar 2016  
  
  %\cite{Barranco:2005yy}
\bibitem{Barranco:2005yy} 
  J.~Barranco, O.~G.~Miranda and T.~I.~Rashba,
  %``Probing new physics with coherent neutrino scattering off nuclei,''
  JHEP {\bf 0512}, 021 (2005)
  %doi:10.1088/1126-6708/2005/12/021
  [hep-ph/0508299].
  %%CITATION = doi:10.1088/1126-6708/2005/12/021;%%
  %64 citations counted in INSPIRE as of 16 Mar 2016
  
  %\cite{Dutta:2015vwa}
\bibitem{Dutta:2015vwa} 
  B.~Dutta, R.~Mahapatra, L.~E.~Strigari and J.~W.~Walker,
  %``Sensitivity to $Z$-prime and nonstandard neutrino interactions from ultralow threshold neutrino-nucleus coherent scattering,''
  Phys.\ Rev.\ D {\bf 93}, no. 1, 013015 (2016)
  %doi:10.1103/PhysRevD.93.013015
  [arXiv:1508.07981 [hep-ph]].
  %%CITATION = doi:10.1103/PhysRevD.93.013015;%%
  %5 citations counted in INSPIRE as of 16 Mar 2016
      
%\cite{Mirabolfathi:2015pha}
\bibitem{Mirabolfathi:2015pha} 
  N.~Mirabolfathi, H.~R.~Harris, R.~Mahapatra, K.~Sundqvist, A.~Jastram, B.~Serfass, D.~Faiez and B.~Sadoulet,
  %``Toward Single Electron Resolution Phonon Mediated Ionization Detectors,''
  arXiv:1510.00999 [physics.ins-det].
  %%CITATION = ARXIV:1510.00999;%%
  %2 citations counted in INSPIRE as of 14 Mar 2016          
      
  %\cite{Andringa:2015tza}
\bibitem{Andringa:2015tza} 
  S.~Andringa {\it et al.} [SNO+ Collaboration],
  %``Current Status and Future Prospects of the SNO+ Experiment,''
  Adv.\ High Energy Phys.\  {\bf 2016}, 6194250
  %doi:10.1155/2016/6194250
  [arXiv:1508.05759 [physics.ins-det]].
  %%CITATION = doi:10.1155/2016/6194250;%%
  %3 citations counted in INSPIRE as of 17 Mar 2016
 
  %\cite{Hochberg:2015fth}
\bibitem{Hochberg:2015fth} 
  Y.~Hochberg, M.~Pyle, Y.~Zhao and K.~M.~Zurek,
  %``Detecting Superlight Dark Matter with Fermi-Degenerate Materials,''
  arXiv:1512.04533 [hep-ph].
  %%CITATION = ARXIV:1512.04533;%%
  %1 citations counted in INSPIRE as of 16 Mar 2016  
    
\end{thebibliography}
\end{document}